\documentclass{emulateapj}

\newcommand{\ebv}{$E(B-V)$}

\newcommand{\Msun}{ $M_{\odot}$}

\newcommand{\zs}{$z$$\sim$}
\newcommand{\zeq}{$z$$=$}
\newcommand{\sqam}{arcmin$^2$}

\newcommand{\inv}{$^{-1}$}

\newcommand{\UGRI}{$U_n G {\cal R} I$}

\newcommand{\Rlim}{${\cal R}_{lim}$}

\newcommand{\lbg}{Lyman Break Galaxy}

\newcommand{\s}{$\sim$}

\newcommand{\Mstar}{$M^*$}
\newcommand{\Lstar}{$L^*$}
\newcommand{\Mchar}{$M^*_{LBG}$}
\newcommand{\Lchar}{$L^*_{LBG}$}
\newcommand{\Mlim}{$M_{lim}$}
\newcommand{\Llim}{$L_{lim}$}

\newcommand{\chisq}{$\chi^2$}
\newcommand{\Muv}{$M_{1700}$}
\newcommand{\Veff}{$V_{eff}$}

\newcommand{\rootN}{$\sqrt{N}$}

\newcommand{\fesc}{$f_{esc}^{rel}$}

\newcommand{\kdfdata}{KDF~I}
\newcommand{\kdflf}{KDF~II}

\shorttitle{Keck Deep Fields. III.}
\shortauthors{Sawicki \& Thompson}

\begin{document}


\title{Keck Deep Fields. III. Luminosity-dependent Evolution of the 
Ultraviolet Luminosity and Star Formation Rate Densities at \zs4, 3, 
and 2\altaffilmark{1}}

\author{Marcin Sawicki} 
\affil{
Department of Physics, University of California, Santa Barbara, CA
93106, USA; and Dominion Astrophysical Observatory, Herzberg Institute
of Astrophysics, National Research Council, 5071~West Saanich Road,
Victoria, B.C., V9E 2E7, Canada }

\email{sawicki@physics.ucsb.edu}

\author{David Thompson} 
\affil{
Large Binocular Telescope Observatory, 
University of Arizona, 
933 N.\ Cherry Ave., 
Tucson, AZ 85721-0065, USA; and 
Caltech Optical Observatories, 
California Institute of Technology,
MS 320-47, Pasadena, CA 91125, 
USA
}
\email{dthompson@as.arizona.edu}

\slugcomment{Accepted for publication in the Astrophysical Journal}

\altaffiltext{1}{Based on data obtained at the 
W.M.\ Keck Observatory, which is operated as a scientific partnership
among the California Institute of Technology, the University of
California, and NASA, and was made possible by the generous financial
support of the W.M.\ Keck Foundation.}

\begin{abstract}
We use our very deep Keck Deep Fields \UGRI\ catalog of \zs4, 3, and 2
UV-selected star-forming galaxies to study the evolution of the
rest-frame 1700\AA\ luminosity density at high redshift --- a study
that is motivated by our finding of luminosity-dependent evolution of
the galaxy luminosity function at high redshift.  Ours is the most
robust UV luminosity density measurement to date at these redshifts as
it uses a well-tested object selection technique, several independent
sightlines, and probes deep into the galaxy luminosity function.  The
ability to reliably constrain the contribution of faint galaxies is
critical and our data do so as they reach deep into the galaxy
population, to \Mchar+2 even at \zs4 and deeper still at lower
redshifts (where \Mchar=$-$21.0 is the Schechter function's
characteristic magnitude for Lyman break galaxies (LBGs) and
\Lchar\ is the corresponding luminosity).  We find that the luminosity 
density at high redshift is dominated by the hitherto poorly studied
galaxies fainter than \Lchar, and, indeed, the the bulk of the UV
light in the high-$z$ Universe comes from galaxies in the rather
narrow luminosity range $L$=0.1--1\Lchar.  It is these faint galaxies
that govern the behavior of the {\it total} UV luminosity density.
Overall, there is a gradual rise in luminosity density starting at
\zs4 or earlier (we find twice as much UV light at \zs3 as at \zs4),
followed by a shallow peak or a plateau within \zs3--1, and then
followed by the well-know plunge at lower redshifts.  Within this
total picture, luminosity density in sub-\Lchar\ galaxies evolves more
rapidly at high redshift, $z$$\gtrsim$2, than that in more luminous
objects. However, this is reversed at lower redshifts, $z$$\lesssim$1,
a reversal that is reminiscent of galaxy downsizing, albeit here thus
far seen only in galaxy luminosity and not yet in galaxy mass.  Within
the context of the models commonly used in the observational
literature, there seemingly aren't enough faint {\it or} bright LBGs
to maintain ionization of intergalactic gas even as late as \zs4.
This is particularly true at earlier epochs and even more so if the
faint-end evolutionary trends we observe at \zs3 and 4 continue to
higher redshifts.  Apparently the Universe must be easier to reionize
than some recent studies have assumed.  Nevertheless, sub-\Lchar\
galaxies do dominate the total UV luminosity density at $z$$\gtrsim$2
and this dominance further highlights the need for follow-up studies
that will teach us more about these very numerous but thus far largely
unexplored systems.

\end{abstract}


\section{INTRODUCTION}

The global, volume-averaged UV luminosity density of the Universe, and
the global star formation rate density that is often derived from it,
contain important information regarding the formation and evolution of
galaxies, the populating of the Universe with stars, and the
production of heavy elements in the cosmological context.

The first measurement of the time evolution of the cosmic UV
luminosity density was done by Lilly et al.\ (1996) who found a
remarkable drop in the luminosity density over the redshift interval
from \zs1 to \zs0 which, with the help of stellar population synthesis
models, they interpreted as an order-of-magnitude decline in cosmic
star formation rate since
\zs1.  The public release of Hubble Deep Field data soon thereafter (HDF;
Williams et al.\ 1996) made it possible to extend such studies to
higher redshifts, \zs1--4, and in nearly concurrent work (the two
papers were published only a month apart) Madau et al.\ (1996) placed
lower limits\footnote{The Madau et al.\ (1996) analysis included
neither incompleteness corrections nor a uniform limiting magnitude.}
on the star formation rate density at \zs3 and 4, while Sawicki, Lin,
\& Yee (1997) estimated the UV luminosity density in
$M_{UV}(AB)$$>$$-$15 galaxies over \zs1.5--3.5, and concluded that the
onset of star formation in the Universe occurred at $z$$>$3.5.

Much subsequent work has focused on either extending the analysis to
higher redshifts (e.g., Lehnert \& Bremer 2003; Bunker et al.\ 2004;
Yan et al.\ 2004; Bouwens et al.\ 2006) in attempts to identify the
galaxy population responsible for cosmic reionization, or on assessing
the corrections for dust either by measuring attenuation in
optically-selected galaxies (e.g., Sawicki \& Yee 1998; Papovich et
al.\ 2001; Shapley et al.\ 2001; Reddy \& Steidel 2004; Iwata, Inoue,
\& Burgarella 2005a), tracking down the contribution of rare but
vigorously star-forming sub-millimeter-selected galaxies (e.g., Smail
et al.\ 1997; Barger et al.\ 1998; Eales et al.\ 1999; Chapman et al.\
2005; Sawicki \& Webb 2005), or attempting to relate these two
star-forming populations (e.g., Chapman et al. 2000; Sawicki 2001;
Baker et al. 2001, Webb et al. 2003).

However, the early HDF-based results (e.g., Madau et al.\ 1996, 1998;
Sawicki et al.\ 1997; Connolly et al.\ 1997) on which the picture at
\zs1--4 was based suffered from a serious drawback: they were based on
a single, tiny field and thus subject to both Poisson noise and
potential systematic problems related to large-scale structure effects
(often referred to as ``cosmic variance'').  Recognizing this as a
serious problem, Steidel et al.\ (1999) used their large,
spectroscopically-calibrated and multi-field Lyman break galaxy
samples to constrain the UV luminosity function and cosmic star
formation rate density at \zs3 and 4.  Their conclusion was that there
was no evidence for evolution in either the galaxy luminosity function
or cosmic SFR density from \zs3 to \zs4.  However, the Steidel et al.\
ground-based surveys contain only galaxies brighter than \Mstar+1 at
\zs3 and \Mstar at \zs4.  Consequently,  while the bright end of the 
galaxy population in their study is reasonably well constrained, these
authors, too, had to rely on HDF data to measure the LF below
$\sim$\Lstar.  Specifically, at \zs3 they used the HDF LBG counts to
constrain the LF's faint end, reporting it to be $\alpha=-1.6$;
however, lacking sufficient statistics for faint \zs4 galaxies in the
HDF they were then forced to {\it assume} the same $\alpha=-1.6$ faint
end slope for their \zs4 measurement.  Thus, the Steidel et al.\
(1999) \zs3 luminosity density measurement is in fact highly reliant
on the single tiny HDF, while their \zs4 luminosity density
measurement, and the conclusion that the SFR density does not evolve
between \zs4 and \zs3 is based on the {\it assumption} of a
non-evolving faint end of the LF.

\begin{figure}
\includegraphics[width=8cm]{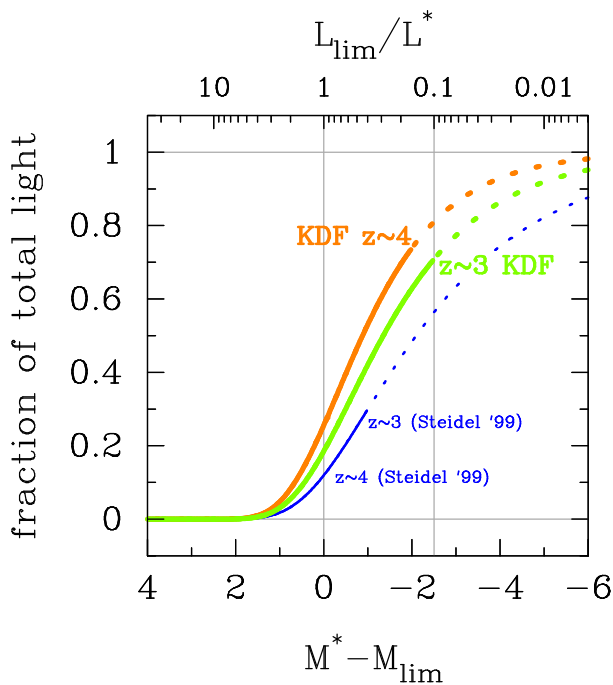}
\caption{\label{fractotallight.fig} 
The amount of total light above a limiting magnitude \Mlim in a
Shechter function with faint end slope $\alpha$.  The colored lines
represent our KDF results at \zs3 (green) and \zs4 (orange) and the
ground-based \zs3 and \zs4 surveys of Steidel et al.\ (1999; blue).
The solid parts of these lines show the directly observed galaxies
(down to the respective limiting depths of these surveys) and the
dashed lines illustrate extrapolations below the completeness limits.
The Steidel et al.\ (1999) ground-based work detects only $\sim$1/3 of
the UV light at \zs3 (and even less at \zs4), while the KDF captures
$\sim$75\% of the UV light at these redshifts.}
\end{figure}

Meanwhile, the contribution of faint, sub-\Lstar galaxies to the
cosmic luminosity density and star formation rate cannot be stressed
strongly enough.  As Fig.~\ref{fractotallight.fig} illustrates, for
most reasonable luminosity functions it is exactly the faint, hitherto
poorly constrained sub-\Lstar\ galaxies that produce the bulk of the
total light.  The Steidel et al.\ (1999, 2003) ground-based surveys
are so shallow that they capture only about a third of the total light
at \zs3 and less than $\sim$20\% at \zs4 and similar limitations apply
to much of the recent work at higher redshifts.  The inference about
the remaining yet dominant bulk of the UV luminosity density is based
on poorly-constrained HDF-based extrapolations to fainter
luminosities.

Recognizing the limitations of the Steidel et al.\ (1999, 2003, 2004)
surveys, we have recently undertaken an imaging survey specifically
targeting intrinsically faint, sub-\Lstar\ Lyman break galaxies.  Our
survey, the Keck Deep Fields (KDF; Sawicki \& Thompson 2005, 2006)
selects \zs4, 3, and 2 galaxies using the {\it very same} \UGRI\
filter set and color-color selection criteria as those used by the
Steidel team and, indeed, our photometry is even further calibrated
onto that of Steidel et al.\ since some of our KDF fields are
co-located with theirs.  This approach gives us a uniquely
well-defined, characterized, and robust sample, but one that probes
1.5 magnitudes deeper than the Steidel et al.\ work and so reaches
significantly below \Lstar: even at \zs4 we reach \Mstar+2 and so can
constrain the shape of the all-important faint end of the luminosity
function very well.

Using these very deep KDF data, we found a shallower LF faint-end
slope at \zs3 than did Steidel et al.\ (1999): $\alpha=-1.43$ vs.\
their HDF-based $\alpha=-1.6$; and an even shallower one at \zs4:
$\alpha=-1.26$ vs.\ their assumed $\alpha=-1.6$ (Sawicki \& Thompson
2006).  Because of our Steidel-like sample selection combined with an
unprecedented wide range of tests that we carried out on our sample,
our LF measurements are uniquely robust against systematic effects.
Details of the KDF observations and LF measurements are given in
Sawicki \& Thompson (2005) and Sawicki \& Thompson (2006),
respectively, and are briefly revied in \S~\ref{data.sec}.  The
purpose of the present paper is to explore how these new, more
accurate, shallower LFs impact our view of the cosmic UV luminosity
density and star formation history of the Universe, by taking proper
account of the evolving number density of faint, sub-\Lstar galaxies.

As in all the papers in the KDF series, in the present paper we use
the AB flux normalization (Oke, 1974) and adopt $\Omega_M$=0.3,
$\Omega_{\Lambda}$=0.7, and $H_0$=70~km s\inv Mpc\inv.  We also define
the characteristic magnitude \Mchar$\equiv$$-$21.0, and a
corresponding characteristic luminosity \Lchar. These paramaters
correspond very closely to the value of \Mstar\ at \zs3 and \zs4 as
measured by us in the KDF (Sawicki \& Thompson 2006) and at \zs3 by
Steidel et al.\ (1999).

\section{THE KDF GALAXY SAMPLE AND LUMINOSITY FUNCTIONS}\label{data.sec}

In this section we briefly summarize some of the most important
features of the KDF data on which we base our analysis.  A detailed
description of the Keck Deep Field observations, data reductions, and
high-$z$ galaxy selection can be found in Sawicki \& Thompson (2005),
while our \zs4, 3, and 2 luminosity function measurements are
presented and discussed in detail in in Sawicki \& Thompson (2006).

\subsection{The KDF survey}

Our analysis is based on results from our very deep multicolor imaging
survey carried out using a total of 71 hours of integration on the
Keck~I telescope.  These Keck Deep Fields use the same
\UGRI\ filter set and color-color selection techniques as are
used for brighter LBGs in the work of Steidel et al.\ (1999, 2003,
2004).  In contrast to the Steidel et al.\ work, however, the KDFs
reach \Rlim=27; this is 1.5 magnitudes deeper than Steidel et al.\
surveys and significantly below \Lstar\ at $z$=2--4: even at \zs4 we
reach two magnitudes fainter than \Mstar.  Furthermore, the KDFs have
an area of 169 arcmin$^2$ divided into 3 spatially-independent patches
on the sky --- an arrangement that allows us to monitor the effects of
cosmic variance on our results.

An important feature of the KDF is that our use of the \UGRI\ filter
set lets us select high-$z$ galaxies using the color-color selection
criteria defined and spectroscopically tested by Steidel et al.\
(1999, 2003, 2004).  We can thus confidently select sub-$L^*$
star-forming galaxies at high redshift without the need for
spectroscopic characterization of the sample --- characterization that
would be extremely expensive at the faint magnitudes that interest us.
Moreover, this commonality with the Steidel et al.\ surveys has
allowed us to reliably combine our data with theirs, thereby for the
first time consistently spanning a large range in galaxy luminosity at
high redshift.

To \Rlim=27, the KDF contain 427, 1481, 2417, and 2043
\UGRI-selected star-forming galaxies at $z$$\sim$4,
$z$$\sim$3, $z$$\sim$2.2, and $z$$\sim$1.7, respectively, selected
using the Steidel et al.\ (1999, 2003, 2004) color-color selection
criteria.  At our completeness limit, the KDF data probe galaxies with
UV luminosities that correspond to star formation rates (uncorrected
for dust obscuration) of only 1--2\Msun/yr.

\subsection{The \zs2, 3, and 4 luminosity functions}\label{theLFs.sec}

Following Steidel et al.\ (1999), our LF calculation uses the
effective volume technique, computing \Veff\ via recovery tests of
artificial high-$z$ galaxies implanted into the images.  To increase
the luminosity range over which we can robustly study the high-$z$
LFs, at the bright end of the \zs3 and \zs4 samples we augmented our
KDF data with the results of Steidel et al.\ (1999).  Because of the
identical filter set, object selection, and LF measurement technique,
we can do this with little fear of introducing systematic offsets as
can happen when other deep but relatively small-area surveys are
augmented with shallower, wide-field data.

\begin{figure}
\includegraphics[width=8cm]{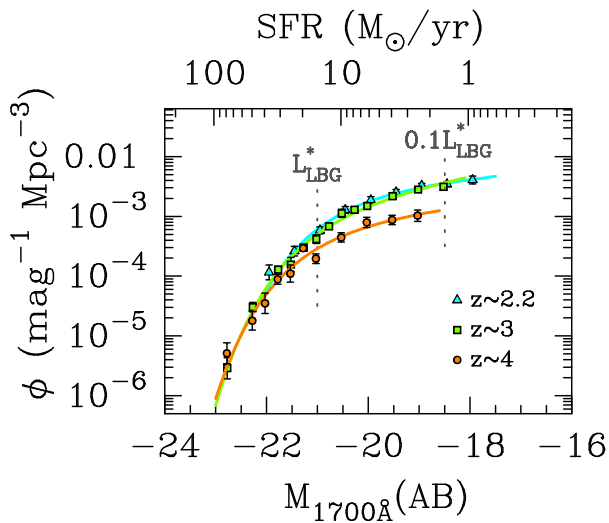}
\caption{\label{LF.fig} The luminosity functions of UV-selected 
star-forming galaxies at \zs4, 3, and 2.2 derived by Sawicki \&
Thompson (2006). The \zs4 and \zs3 LFs are very robust against
systematic effects; at \zs2.2, the number density of galaxies is {\it
not less} than that shown and if it is higher than shown, then it is
so by no more than a factor of $\sim$2.  }
\end{figure}

In Fig.~\ref{LF.fig} we reproduce the \zs4, 3, and 2.2 luminosity
functions from Sawicki \& Thompson (2006).  In that paper we have
examined a wide range of potential sources of systematic problems,
including effects due to modeling of survey volume, differential
sample selection, field-to-field variance, uncertainties in
$k$-corrections, and so on. On the basis of these tests we have a very
good understanding of the reliability of our LF results.  In summary,
we found that at \zs4 and \zs3, our LF measurements are immune to a
range of potential systematic effects, while our \zs2.2 and \zs1.7 LFs
are more uncertain.  The \zs1.7 LF is likely quite strongly affected
by systematics and so we do not include it in the analysis presented
here.  The $z$$\sim$2.2 LF measurement is {\it possibly} also affected
by systematics but to a much smaller degree than the \zs1.7 one: we
can confidently state that the number density of \zs2.2 galaxies is
{\it not lower} than that shown in Fig.~\ref{LF.fig}, and that if it
is higher, it is higher by no more than a factor of $\sim$2.  The
potential systematic error at \zs2.2 is due to our uncertainty
regarding the survey volume which itself stems from our uncertainty in
the model colors of \zs2.2 galaxies.  The LF shown in
Fig.~\ref{LF.fig} is based on our best and most realistic assumptions,
namely that there is moderately strong dust obscuration of
\ebv=0.15. The potentially twice-higher normalization would ensue in
the extreme and rather unrealistic case of no dust (see Sawicki \&
Thompson 2006 for details) --- a scenario that appears highly unlikely
in view of recent spectral energy distribution studies (Shapley et
al.\ 2005; Sawicki et al.\ 2006).  It must be stressed again, however,
that the $z$$\sim$3 and $z$$\sim$4 LFs, are {\it very} robust against
such and other systematics.

The uncertainties in Fig.~\ref{LF.fig} include both \rootN\
statistical uncertainty in galaxy numbers and a measure of the effects
of large scale structure (``cosmic variance'') obtained from a
bootstrap resampling analysis of the multiple fields in the KDF.  Both
these sources of uncertainty are propagated into our error estimates
for the luminosity density.  We also note that at both \zs4 and \zs3,
we find \Mstar$\approx-21.0$, in good agreement with the
\zs3 Steidel et al.\ 1999 value converted into our cosmology. For
convenience, we have thus defined a reference characteristic magnitude
\Mchar$\equiv-21.0$ and the equivalent characteristic luminosity
\Lchar; these values correspond to a galaxy with a (dust-free) star
formation rate of 15\Msun/yr (e.g., Kennicut 1998).

In contrast to the Steidel et al.\ (1999) results, however, we find a
somewhat shallower faint-end slope $\alpha$.  At \zs3 we find
$\alpha=-1.43$, compared to their $\alpha=-1.6$, and at \zs4 we find
$\alpha$$=$$-1.26$.  This evolving faint-end slope marks one of the
most striking findings of the KDF to date: a differential,
luminosity-dependent evolution of the \lbg\ luminosity function.
Simply put, while the number density of luminous galaxies
($L$$>$\Lstar) remains fixed, the number density of sub-\Lstar\
objects more than doubles from \zs4 to \zs3.  Our numerous tests show
that this result is highly robust and insensitive to systematic biases
(see Sawicki \& Thompson 2006).  In the present paper we turn to
investigate how this differentially evolving galaxy population,
reflected in an evolving luminosity function, impacts our view of the
cosmic luminosity density and star formation rate density.

\section{THE LUMINOSITY DENSITY}\label{lumdens.sec}

The luminosity density is the sum of all the light at a given
wavelength, or range of wavelengths, from all galaxies in a unit
volume.  It can be computed as the luminosity-weighted integral of the
galaxy luminosity function $\phi (L)$:
\begin{equation}\label{deflumdens.eq}
\rho_L = \int^{\infty}_{0} L \phi(L) dL, 
\end{equation}
which, for a luminosity function of the Schechter (1976) form,
$\phi(L) dL = \phi^* (L/L^*)^{\alpha} e^{-L/L^*} dL/L$, can be
expressed in terms of the gamma function, $\Gamma(x)=\int_0^\infty
e^{-t} t^{x-1} dt$, as
\begin{equation}\label{lumdensasgammafunction.eq}
\rho_L = \phi^* L^* \Gamma(\alpha+2).
\end{equation}

Because of the difficulty of constraining the number density of faint
galaxies at high redshift, most studies focus on the partial, or
incomplete, luminosity density --- i.e., the luminosity density due to
objects brighter than some limiting luminosity \Llim.  We, too, will
do so here at first by focusing on galaxies brighter than 0.1\Lchar\
(\S~\ref{brightlumdens.sec}) before presenting the {\it total}
luminosity density (\S~\ref{totallumdens.sec}) and then examining the
contributions to that total that are made by galaxies with
luminosities in various luminosity bins (\S~\ref{lumdensbylum.sec}).

\subsection{Luminosity in galaxies brighter than 0.1\Lchar}
\label{brightlumdens.sec}

Given a galaxy luminosity function $\phi(L)$, the luminosity density
due to galaxies brighter than \Llim\ is
\begin{equation}
\label{lumdensincomplete.eq}
\rho_L(L>L_{min}) = \int^{\infty}_{L_{min}} L \phi(L) dL. 
\end{equation}
For the Schechter LF parametrization this can be expanded as
\begin{equation}\label{partint1.eq}
\rho_L(L>L_{min}) 
= \phi^* L^* [ \Gamma(\alpha+2) - \gamma(\alpha+2, L_{min}/L^*)], 
\end{equation}
where $\gamma(x,l)=\int_0^l e^{-t} t^{x-1} dt$ is the lower incomplete
gamma function (e.g., Arfken 1985, Press et al.\ 1986).

We adopt \Llim=0.1\Lchar, which corresponds to \Muv=$-$18.5 or, in the
absence of dust, a galaxy with a star formation rate of
$\sim$1.5\Msun/yr (Kennicutt 1998).  As Fig.~\ref{fractotallight.fig}
shows, this choice of \Llim\ captures approximately 75\% of the UV
light at \zs4 and 3 for the luminosity functions we measure in the
KDF.

\begin{figure}
\includegraphics[width=8cm]{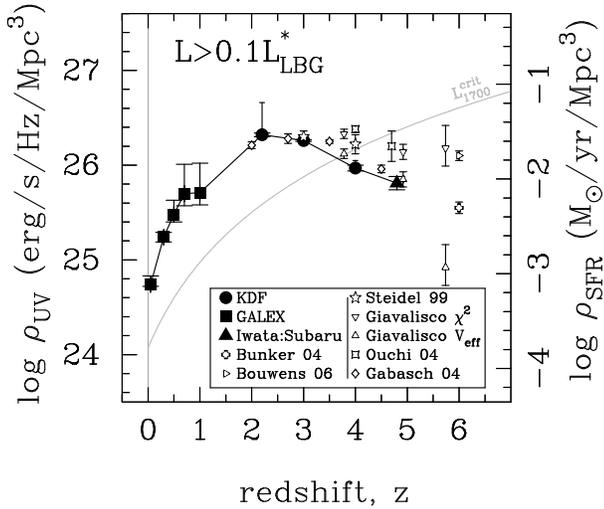}
\caption{\label{lumdens-bright.fig}  
The UV luminosity density in galaxies brighter than 0.1\Lchar.  Solid
symbols show values from the present KDF work at \zs2, \zs3, and \zs4,
from GALEX (at $z$$\leq$1) and from Iwata et al.\ (2006; \zs5).  Also
shown are results from other recent work.  Two errorbars are shown for
the KDF points: the (generally smaller) errorbars with long terminals
include both Poisson statistics and a bootstrap-resampling estimate of
cosmic variance (many of the other surveys show Poisson errorbars
only); the (generally larger) errorbars, most prominent at \zs2,
include an estimate of the {\it maximum} possible systematic error
effect in our modelling of the survey volume (see Sawicki \& Thompson
2006 for details).  The star formation rate density (right axis) is
calculated from the UV luminosity density without correction for dust.
The curve labelled $L_{1700}^{crit}$ shows the amount of 1700\AA\
light expected --- on the basis of the calculation in
\S~\ref{ionization} --- to be needed if starforming galaxies are to
maintain ionization of the intergalactic gas.  }
\end{figure}

We show the $\rho_L(L$$>$0.1\Lchar) KDF values with filled circles in
Fig.~\ref{lumdens-bright.fig}, together with GALEX values at lower
redshift (filled squares) and Iwata et al.\ (2006a, 2006b)
values at \zs5 (filled triangle).  Tables~\ref{lumdens.tab} and
\ref{lumdensBM.tab} list the values of luminosity density for the KDF
and for other surveys. The KDF values were computed by applying
Eq.~\ref{lumdensincomplete.eq} to the Schechter function parameters of
Sawicki \& Thompson (2006). The uncertainties associated with our KDF
points in Fig.~\ref{lumdens-bright.fig} (and in all subsequent
figures) include the contributions of both Poisson statistics and
cosmic variance, the latter estimated through bootstrap resampling of
the five KDF fields. Also shown in Fig.~\ref{lumdens-bright.fig} are
the $z$$\leq$1 GALEX UV luminosity density values of Schiminovich et
al.\ (2005) which we have translated from their 1500\AA\ to 1700\AA\
and adjusted to represent only galaxies brighter than 0.1\Lchar\ by
scaling their total luminosity densities using our
Equations~\ref{lumdensincomplete.eq} and \ref{deflumdens.eq} together
with the GALEX LF Schechter parameters given by Arnouts et al.\ (2005)
and Wyder et al.\ (2005).  The \zs5 point is based on new work by
Iwata et al.\ (2006a, 2006b), who select \zs5 LBGs from Subaru
Suprimecam data (the median redshift of their sample is $z$=4.8, but
for simplicity we refer to it here as \zs5).  These new \zs5 results
build on earlier work by these authors (Iwata et al.\ 2003) and are
the deepest and widest-area LF results presently available at \zs5, as
they are drawn from a total of 1290 \sqam\ and reach \Mchar+0.95, or
0.7 mag deeper than the \zs5 LFs of Ouchi et al.\ (2004).  I.\ Iwata
and his collaborators have kindly provided us with their new, deep
estimates of \zs5 incompleteness-corrected LBG number densities which
they have computed using the \Veff\ approach similar to the one we
used in the KDF (see Iwata et al.\ 2003 and I.\ Iwata in preparation
for details).  We then used these number densities to calculate the
Schechter function parameters and their uncertainties following the
same procedure as we used for our lower-$z$ KDF samples (see \kdflf).
We find $M^*_{1700}=-20.80^{+0.40}_{-0.40}$, $\log
\phi^* = -3.10^{+0.19}_{-0.30}$, and $\alpha = -1.01^{+0.50}_{-0.49}$ 
at \zs5 from the Iwata et al.\ data while the corresponding \zs5 UV
luminosity densities are given in Tables~\ref{lumdens.tab} and
\ref{lumdensBM.tab}.

Several interesting trends are apparent in
Fig.~\ref{lumdens-bright.fig}.  First, it is clear that the luminosity
density at \zs4 is substantially lower than that at \zs3.  This is a
direct consequence of the shallower faint-end of the luminosity
function that we find at \zs4.  This drop in luminosity density at
\zs4 is in contrast to the \zs4 result of Steidel et al.\ (1999; small
star symbols), but this is because these authors {\it assumed} a
steeper, non-evolving $\alpha=-1.6$ faint-end slope at this redshift.
Our KDF results, instead, show an evolutionary trend whereby the
luminosity density increases with time from \zs4 to \zs3.  As we show
in more detail in
\S~\ref{lumdensbylum.sec}, this trend is due to the evolution of the
population of sub-\Lchar\ galaxies.  It is not clear whether this
evolutionary trend continues to \zs2, or whether there is a plateau in
the luminosity density at \zs3--1.  Our data favor a plateau, but a
peak at \zs2 cannot be fully ruled out due to potential systematic
problems at that redshift (see \S~\ref{theLFs.sec} and Sawicki \&
Thompson 2006 for details).
 
It also appears plausible that the evolution we see at
\zs4$\rightarrow$3 in the KDF data may have started at earlier times:
if we consider the Iwata et al.\ (2006) \zs5 result then there appears
to be a monotonic increase in the luminosity density produced by
galaxies brighter than 0.1\Lchar\ that extends from very early times,
at least \zs5, until at least \zs3.

The GALEX points illustrate the now well-known drop in luminosity
density from \zs1 to \zs0. Together with our KDF results at
\zs4, 3, and 2, and the Iwata et al.\ results at \zs5, the following
trend is clear: the luminosity density, at least in galaxies brighter
than 0.1\Lchar, increases steadily from the earliest redshifts until
at least \zs3; it then culminates somewhere between \zs3 and \zs1 as
either a peak or a broad plateau; and it then drops precipitously from
\zs1 to \zs0.  This picture is consistent with that first presented by
Sawicki et al.\ (1997), who found a broad peak in the luminosity
density at \zs2.5 and an onset of cosmic star formation at
$z$$>$3.5\footnote{We note that Madau et al.\ 1996 presented only
lower limits on the high-redshift luminosity densities and that these
values were corrected for incompleteness only later in Madau et al.\
1998.}.

\subsection{The total luminosity density}\label{totallumdens.sec}

Our analysis thus far (summarized in Fig.~\ref{lumdens-bright.fig}),
has concerned itself with luminosity density due to galaxies brighter
than 0.1\Lchar\ only.  This is an approach common in the literature,
but one that does not present a complete census of the cosmic
luminosity density.  We thus now turn to the question of how our
results change if we incorporate the contribution of extremely faint
galaxies, $L$$<$\Lchar\ --- galaxies that are too faint to be seen in
{\it any} of the surveys to date at high redshift, including ours.

The {\it total} luminosity density is obtained by integrating the
luminosity function over {\it all} luminosities from 0 to $\infty$
(Eq.~\ref{deflumdens.eq}). This calculation involves an extrapolation
to magnitudes fainter than our limiting magnitudes, but the depth of
our KDF data allows this to be a robust extrapolation: (1) because of
their large area and depth the KDFs have constrained the LFs'
faint-end slopes sufficiently well, and, (2) the depth of our data is
such that the extrapolation involves only $\sim$25\% of the total
light (Fig.~\ref{fractotallight.fig}).  Similar remarks apply to the
surveys of Iwata et al.\ (2006) at \zs5 and GALEX at $z$$\lesssim$1.
Surveys with smaller areas and/or shallower limiting magnitudes must
make much bigger and more uncertain corrections, but this is not the
case for the KDF.

\begin{figure}
\includegraphics[width=8cm]{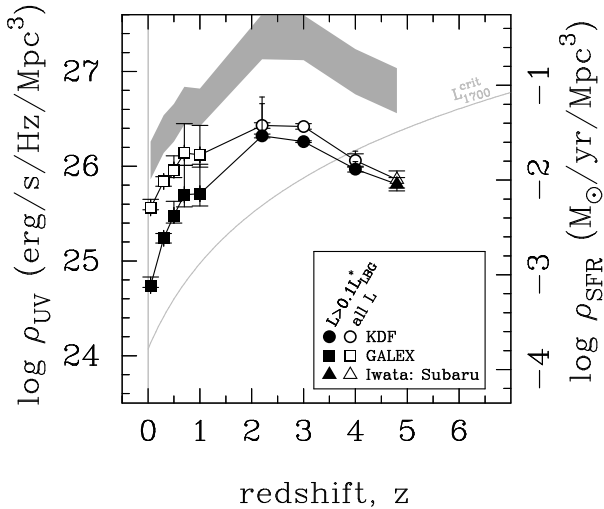}
\caption{\label{lumdens-wdust.fig}  
The {\it total} luminosity density and dust corrections.  Filled
symbols show the luminosity density due to galaxies brighter than
\Lchar, while open symbols show the total (all $L$) luminosity
density.  The gray band shows the allowable range after the light from
the total (all $L$) UV-bright galaxy population has been corrected for
dust obscuration as discussed in \S~\ref{totallumdens.sec}.  }
\end{figure}

The {\it total} luminosity densities are shown in
Fig.~\ref{lumdens-wdust.fig} with open symbols and are compared with
$L$$>$0.1\Lchar\ results of \S~\ref{brightlumdens.sec} that are
represented by filled points.  The difference between the
$L$$>$0.1\Lchar\ and the total results is very small at high redshift,
$z$$\geq$2.  This is so because the faint-end slopes measured by us in
the KDF (and by Iwata et al.\ 2006 at \zs5) are relatively shallow,
resulting in relatively little additional light in galaxies fainter
than 0.1\Lchar.  However, although the effect is small here, the
importance of reliably measuring the faint end slope of the LF cannot
be understated: had the faint-end slopes turned out to be
significantly steeper, much stronger differences would have ensued. As
it is, most of the light at high redshift resides in galaxies with
$L$=0.1--1\Lchar\ (as we show in \S~\ref{lumdensbylum.sec}),
underscoring the importance of accurately measuring the LF's faint end
slope at luminosities fainter than \Lchar.  At lower redshifts,
$z$$\leq$1, there is a fairly significant difference between the total
and the $L$$>$0.1\Lchar\ values. However, this is not because the
low-$z$ faint-end slope is very steep (it isn't), but primarily
because at $z$$\leq$1 \Lstar\ is a magnitude or more fainter than our
fixed high-$z$ \Lchar. This difference ensures that the 0.1\Lchar\
limit of \S~\ref{brightlumdens.sec} does not encompass most of the
low-$z$ light.

\begin{figure}
\includegraphics[width=8cm]{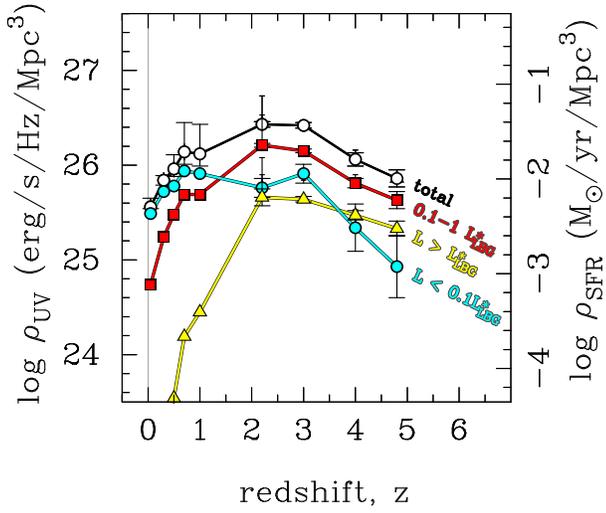}
\caption{\label{lumdens.fig}  
Contributions of galaxies with different individual luminosities to
the total UV luminosity density of the Universe.  Open circles show
the total luminosity density while colored symbols show the luminosity
density split by galaxy luminosity.  The split is at \Lchar\ and
0.1\Lchar (i.e., \Muv=$-21.0$ and $-18.5$, respectively). Galaxies
with $L$=0.1-1\Lchar\ (red squares) dominate the total luminosity
density at high redshift, while those fainter than 0.1\Lchar\ take
over at $z$$\lesssim$1.  At no redshift are galaxies brighter than
\Lchar\ dominant.  } \end{figure}

At $z$$\leq$1 the total UV luminosity becomes dominated by very faint
galaxies, galaxies with luminosities below 0.1\Lchar\ (see
Fig.~\ref{lumdens.fig}).  It is curious to note that the total
luminosity density at \zs0.5 is similar to that at \zs4, which is not
the case when the comparison is restricted to galaxies brighter than
0.1\Lchar\ only.  Nevertheless, while the decline from \zs1 to \zs0 is
less precipitous when one considers the {\it total} luminosity density
rather than just that due to galaxies brighter than 0.1\Lchar, the
broad qualitative picture is the same in both cases: a gradual
build-up in the luminosity density from very high redshift which is
followed by a peak or a plateau between \zs3 and \zs1 which in turn is
followed by a decline to \zs0.

This broad qualitative picture is also only mildly affected when one
corrects for the effects of interstellar dust.  Here, at each redshift
we have applied a {\it range} of plausible dust corrections and the
resulting dust-corrected star formation rate densities are shown as a
gray band in Fig.~\ref{lumdens-wdust.fig}.  We corrected the
$z$$\geq$2 values by factors of 5--15 as suggested by various studies
of extinction in Lyman Break Galaxies (e.g., Sawicki et al.\ 1998;
Papovich et al.\ 2001; Shapley et al.\ 2001, 2005; Vijh, Witt, \&
Gordon 2003; Reddy \& Steidel 2004; Iwata et al.\ 2005) and the
$z$$<$2 GALEX values by 2--5 (Schiminovich et al.\ 2005; see also,
e.g., Tresse et al.\ 2002; P\'erez-Gonz\'alez et al.\ 2003). After
applying dust corrections we can now properly regard
Fig.~\ref{lumdens-wdust.fig} as showing us the cosmic star formation
rate history of UV-bright galaxies.

Extremely dust-obscured galaxies such as high-$z$ SCUBA sources (e.g.,
Smail et al.\ 1997; Barger et al.\ 1998; Eales et al.\ 1999; Chapman
et al.\ 2005) are not included in the dust-corrected star formation
rate densities shown in Fig.~\ref{lumdens-wdust.fig}.  Most such
galaxies have UV luminosities that heavily underrepresent their star
formation rates and thus such galaxies are not properly included if
our census is to be regarded as a measurement of star formation
activity in the Universe.  The number density of such heavily obscured
sources at {\it low redshift} is too low to make a large contribution
to the star formation rate density.  However, at high redshift,
\zs2, their numbers are sufficiently large that their contribution is
significant and including them in our census would raise the star
formation rate density shown in Fig.~\ref{lumdens-wdust.fig}.  The
increase would be highest at \zs2 where the redshift distribution of
SCUBA sources appears to peak (Chapman et al.\ 2005). The correction
to the star formation rate density at that epoch may be up to a factor
of two, i.e., 0.3 in dex (Chapman et al.\ 2005; Reddy et al.\ 2005).

Taken together, however, the incorporation of dust corrections --- be
they only dust corrections to the UV-selected galaxies or also the
inclusion of the contribution of heavily-obscured sources --- does not
qualitatively affect the picture shown in
Figs.~\ref{lumdens-bright.fig} and \ref{lumdens-wdust.fig}.  Dust or
no dust, the star formation rate density of the universe apparently
began rising at some high redshift $z$$\gtrsim$4, continued to rise
until at least \zs3, climaxed in a peak or a plateau between \zs3 and
\zs1, and then dropped to \zs0.

While the decline from \zs1 to the present epoch has been well
established for a long time (e.g., Lilly et al.\ 1996; Shiminovich et
al.\ 2005), the behavior of the cosmic star formation rate denstity at
$z$$\gtrsim$3 was less clear: the Steidel et al.\ (1999) picture of a
plateau extending from \zs3 to \zs4 and possibly beyond (e.g.,
Giavalisco et al.\ 2004b; Ouchi et al.\ 2004) was for a long time the
gold standard in the field.  However, the Steidel et al.\ (1999)
no-evolution result was based on the measurement of a non-evolving
LF's bright end and the {\it assumption} of a non-evolving faint-end.
Our KDF results clearly show that, in contrast to these assumptions,
the luminosity function evolves and this evolution is reflected in the
luminosity density which evolves with time at high redshift.

\subsection {Dependence on galaxy luminosity}\label{lumdensbylum.sec}

In the previous Section we have shown that the UV luminosity density
exhibits a rise from early epochs to \zs3, a peak or plateau between
\zs3 and \zs1, and then a decline to \zs0.  We now ask the question: 
which galaxies are responsible for this picture? To address this
question we split our analysis into three luminosity intervals,
computing the luminosity density in galaxies brighter than \Lchar,
galaxies between \Lchar\ and 0.1\Lchar, and those fainter than
0.1\Lchar.

The luminosity density due to just those objects with luminosities in
the range $L_{max}>L>L_{min}$ is given by
\begin{eqnarray}\label{partint2.eq}
\rho_L(L_{max}>L>L_{min}) & = & \int^{L_{max}}_{L_{min}} L \phi(L) dL \\
                          & = & \phi^* L^* [\gamma(\alpha+2,
                          L_{min}/L^*) \nonumber \\ & & - \gamma(\alpha+2,
                          L_{max}/L^*)] \nonumber.
\end{eqnarray}
We make the boundaries of our luminosity intervals at \Lchar\ and at
0.1\Lchar.  Note that to compute the last (faintest) of those three
intervals we must extrapolate the faint-end slope of the LF at all
redshifts, but the data we consider here (the KDF, GALEX, and the \zs5
Subaru data of Iwata et al.\ 2006) all go sufficiently deep to measure
the faint-end slopes well, thus making these extrapolations
acceptable.

In Fig.~\ref{lumdens.fig} we plot the redshift evolution of the
luminosity-dependent luminosity densities (colored symbols and lines)
compared with the {\it total} luminosity density from
Fig.~\ref{lumdens-wdust.fig} in \S~\ref{totallumdens.sec} (open
circles and black line).  At high redshift, $z$$>$2, the luminosity
density is dominated by galaxies in the narrow luminosity range
(0.1--1)\Lchar: galaxies that are brighter than \Lchar\ or those that
are fainter than 0.1\Lchar\ contribute relatively little light to the
total at these redshifts!  In contrast there is little evolution in
the bright end of the galaxy population between \zs2, 3, and 4 (and 5)
and this results in a nearly constant luminosity density of bright,
super-\Lchar\ galaxies over \zs2--5.  To date, most of the focus at
high redshift has been on galaxies around \Lchar.  But it is the
sub-\Lstar\ galaxies that, as we have illustrated here, dominate the
total luminosity density at $z$$\gtrsim$2.  This dominant
sub-population is only now for the first time starting to be robustly
probed by our KDF survey and by follow-up studies that are now either
underway or being planned.

We conclude this section by summarizing our main points. At high
redshift, $z$$>$2, there is a rise in the total luminosity density
with time. This rise is not due to changes in the population of bright
galaxies, $L$$\gtrsim$\Lchar, but is instead directly related to the
rise in the number density of faint, sub-\Lchar\ galaxies and the
steepening of the faint end slope of the luminosity function that
reflects it.

\section{DISCUSSION}

\subsection{Comparison with other work}
\label{comparisons.sec}

In this section we compare our results with those of other surveys.
Figure~\ref{lumdens-bright.fig} shows results from several other
programs that recently studied the cosmic UV luminosity density at
high redshift.  We consider here a number of recent surveys, but ---
with the exception of the work of Steidel et al.\ (1999) --- we avoid
results that are based on the very small area HDFs (e.g., Madau et
al.\ 1996; 1998; Sawicki et al.\ 1997; Connolly et al.\ 1997; Arnouts
et al.\ 2005) because such results have to be held suspect given the
small field of view of the HDFs and the resulting susceptibility to
cosmic variance.  We have, however, included some \zs6 results based
on the small Hubble Ultra Deep Field since there are at present no
other data that probe significantly below \Lchar\ at that redshift.
In all the surveys that we do consider we have converted the published
results to a consistent limiting depth of $L$$=$0.1\Lchar\ and have
adjusted them to rest-frame 1700\AA\ values as necessary.  The
superiority of the KDF results vis-a-vis these other surveys lies in
the combination of the KDF's superb depth, large field of view,
multiple sightlines, well-understood sample selection, and an
extensive range of robustness-gauging tests that we carried out as
part of our LF measurement (Sawicki \& Thompson 2006).

\subsubsection{\zs2 and \zs3}\label{comp-z23.sec}

At \zs2 and \zs3 our KDF results are in very good agreement with other
available measurements, namely the photometric redshift results of
Gabasch et al.\ (2004b) and the LBG work of Steidel et al.\ (1999).

The agreement with Steidel et al.\ (1999) \zs3 results is --- at one
level --- not surprising because the KDF uses the very same filter set
and sample selection as the ground-based Steidel et al.\
surveys. Indeed --- our \zs3 and \zs4 luminosity functions incorporate
Steidel et al.\ (1999) data at the bright end (see Sawicki \& Thompson
2006). On the other hand, however, Steidel et al.\ (1999) used the
Hubble Deep Field to constrain the shape of their luminosity function
below \Mchar+1, thereby incurring the uncertainty of differential
selection between their bright and faint galaxies, in addition to
potential cosmic variance issues due to the HDF's small area.
Although there is a small difference between our and their faint-end
LF slopes at \zs3 (our $\alpha$=$-1.43$ vs.\ their $\alpha$=$-1.6$),
there is quite good agreement between the luminosity densities at
$L$$<$0.1\Lchar.

Our KDF results are also in excellent agreement with the \zs2 and \zs3
FORS Deep Field (FDF) and Great Observatories Origins Deep Survey
South (GOODS-S) results of Gabasch et al.\ (2004b).  The FDF is a
single relatively small ($\sim$ 40 \sqam) deep field with sample
selection (full-blown photometric redshifts) that is rather different
than ours.  Their GOODS-S data uses similar sample selection and area
($\sim$ 50 \sqam) as the FDF but is a magnitude shallower than their
FDF data, reaching only $\sim$\Mchar+1 at \zs3 --- depths comparable
to those reached from the ground by Steidel et al.\ (1999).  Despite
the differences in sample selection and the relatively small area of
their significantly deep data (the FDF), there is very good agreement
between the luminosity densities they find and those that we find in
the KDF.

Overall, all three results (KDF, Gabasch et al.\ 2004b, and Steidel et
al.\ 1999) are highly consistent with each other, albeit ours is the
most robust of the three given the KDF's large area and multiple
sightlines (important for mitigating the effects of cosmic variance)
combined with faint limiting magnitude (vital for constraining the
faint end of the galaxy population), and extensive LF robustness
tests.

\subsubsection{\zs4}

At \zs4 we compare our results with Subaru Deep Survey (SDS) work of
Ouchi et al.\ (2004), Giavalisco et al. (2004b) results based on the
Great Observatories Origins Deep Survey (GOODS; Giavalisco et al.\
2004a) HST data, as well as the aforementioned work of Gabasch et al.\
2004b and that of Steidel et al.\ (1999). At \zs4, the GOODS-S
component of the Gabasch et al.\ (2004b) analysis reaches only
\s\Mchar+0.5, so here we consider their work to be based primarily on
the single-pointing FORS Deep Field.  All the surveys considered use
variants of the Lyman Break color-color selection technique, with the
exception of the FDF, which uses full photometric redshifts.

These four surveys (Ouchi et al. 2004; Gabasch et al.\ 2004b;
Giavalisco et al.\ 2004; and Steidel et al.\ 1999) do not present a
consistent picture, giving a range of luminosity densities that span a
factor of $\sim$2.5.  Let us examine the differences in more detail.

Gabasch et al. (2004b) do not give FDF luminosity density measurements
at \zs4 but present values at \zeq3.5 and 4.5 (see
Fig.~\ref{lumdens-bright.fig}).  However, these values bracket our
\zs4 KDF result and a straightforward interpolation between their 
FDF \zeq3.5 and \zeq4.5 points is in very good agreement with our
data.  This agreement is to be expected given the good agreement
between our LFs and theirs (see Sawicki \& Thompson 2006), but note
that our KDF result is much more robust for reasons outlined in
\S~\ref{comp-z23.sec}.

The results of Steidel et al.\ (1999) at first appear to be somewhat
higher than ours, but this is because of their {\it assumption} that
the faint end of the LF does not evolve from \zs3.  With a shallower
faint-end $\alpha$ at \zs4, as found in the KDF, their results would
be in agreement with ours.

Giavalisco et al.\ (2004b) give two measurements from the GOODS data,
obtained using two different techniques. One measurement (from their
\chisq\ technique) paints a picture of a luminosity density that does
not differ from that at \zs3, while the other (from their
\Veff\ approach) yields an evolving luminosity density, in agreement
with our results.  Giavalisco et al.\ prefer their \chisq\ method and
results, but our KDF data are in much better agreement with their
lower, \Veff\ point.

The SDS work of Ouchi et al. (2004) gives the highest luminosity
density value but this value may well be prejudiced by potential
problems with their measurement of the LF's faint end: at \zs4 their
data probes one magnitude fainter than \Mchar\ and they find an
extremely steep faint-end slope of $\alpha$=$-2.2$ that gives an
infinite {\it total} luminosity density (Eq.~\ref{deflumdens.eq}).  A
strong limitation of their work is that their color selection is not
well calibrated spectroscopically except for a handful of redshifts
and must instead rely heavily on models of galaxy colors.  It is not
clear how strong are the systematic effects inherent in this modeling.

Overall, our results are in very good agreement with those from the
FDF and with the GOODS \Veff\ approach, while we find substantially
less luminosity than does the SDS and the GOODS \chisq\ approach.  We
feel that, given the robustness of our sample selection and LF
measurement, our results are the most reliable of all those currently
available at \zs4.

\subsection{Earlier epochs}

\subsubsection{\zs5}

The literature now contains several measurements of the luminosity
function and cosmic luminosity density at \zs5. Here we discuss those
by Iwata et al (2006) as well as those based on GOODS (Giavalisco et
al.\ 2004b), SDS (Ouchi et al.\ 2004), and FDF (Gabasch et al.\ 2004a,
2004b; note that the FDF results are at \zs4.5).  With the exception
of the Giavalisco et al.\ GOODS results, all these surveys present
both luminosity densities and luminosity functions; of these, the
Iwata et al.\ (2006) Subaru work has the best combination of depth
(0.7 mag deeper than the SDS) and area (an order of magnitude larger
than the FDF).

The Iwata et al.\ (2006), FDF, and the GOODS \Veff\ results all
indicate that the luminosity density at \zs5 is lower than that at
\zs4 in the KDF.  In both the Subaru work of Iwata et al.\ (2006) and
in the FDF (Gabasch et al.\ 2004a) the faint-end slope of the LF is
quite shallow at these redshifts, in apparent continuation of the LF
evolutionary trend that we see in the KDF. It is this shallow
faint-end slope that results in the low luminosity density that these
surveys measure at \zs5.  On the other hand, the GOODS \chisq\ results
as well as those from the SDS would suggest a higher luminosity
density --- one that is almost unchanged from \zs3.  In the SDS work
this is caused by a very steep faint-end slope that they {\it assume}
from their lower-$z$ results (their observations at \zs5 reach only to
\Lchar+0.25).  The \zs3$\rightarrow$4 trend seen in the KDF tends to 
favor a continuing drop in the luminosity density to higher redshifts,
in line with the Iwata et al.\ (2006), FDF, and GOODS-\Veff\ values.

\subsubsection{\zs6}\label{z6.sec}

The situation at \zs6 is tenuous despite much recent effort.  Some
recent results (Bouwens et al.\ 2006; the \chisq\ GOODS results of
Giavalisco et al. 2004b) suggest a \zs6 luminosity density nearly as
high as that at \zs3. Others, (Bunker et al.\ 2004; \Veff\ results of
Giavalisco et al.\ 2004b) give values that are significantly lower
(see Fig.~\ref{lumdens-bright.fig}).

The situation is tenuous for several reasons.  First, at \zs6 sample
selection is generally done using just one color, $i$$-$$z$, which
makes it difficult to estimate redshift distributions, and hence
survey volumes, and hence source number densities.  The absence of a
third, redder band also makes source luminosities {\it very} uncertain
because at \zs6 the $z$-band fluxes that are invariably the starting
point for computing luminosities are attenuated by intergalactic
hydrogen line blanketing. While corrections for this attenuation are
usually applied, such corrections are necessarily statistical because
they rely on an {\it average} rather than exact distribution of
intervening gas, and so are subject to systematic biases that are
difficult to quantify.  Moreover, these corrections are typically
based on the models of Madau (1995); these models are based on
lower-redshift data, and it is unclear how well they work at
\zs6. It is thus not clear how well the fluxes and hence luminosities 
of \zs6 candidates are being estimated.  

Another limitation is that the all-important shape of the faint end of
the \zs6 luminosity function is highly uncertain.  Assuming the
Steidel et al.\ (1999) \zs3 faint-end slope of $\alpha$=$-1.6$ is not
a good approach in light of the fact that the true slope is shallower
even at \zs3 and that it becomes progressively shallower with
increasing redshift (Sawicki \& Thompson 2006; see also Gabasch et
al.\ 2004a).  On the other hand, empirically constraining the number
density of sub-\Lchar\ galaxies from the data is highly problematic
since the only field sufficiently deep for this task is the Hubble
Ultra Deep Field (UDF).  The UDF's small area (12
\sqam) makes it subject to potential cosmic variance effects only
slightly less severe than those affecting the HDFs.  Bouwens et al.\
(2006) argue that cosmic variance effects on UDF scales are not large
at \zs6 (they suggest a 35\% RMS effect), but this seems questionable:
e.g., the difference in comoving luminosity density at redshifts \zs3
and \zs4 between the two $\sim$5 \sqam\ Hubble Deep Fields is a factor
of two (e.g., Ferguson, Dickinson, \& Williams 2000) and it is
reasonable to suppose that similarly large fluctuations can be present
on the just slightly larger scale of the UDF.

In summary, present measurements of the \zs6 UV luminosity density are
highly uncertain. Given that our KDF data show that luminosity density
is dropping {already} from \zs3 to \zs4, it seems unlikely that it
would be as high at \zs6 as it is at \zs3.  A simple extrapolation of
the trend we see between \zs3 and \zs4 in the KDF --- a trend that
seems to continue to \zs5 in the work of Iwata et al.\ 2006 ---
suggests the cosmic UV luminosity density at \zs6 is similar to that
reported by Bunker et al.\ (2004).  To make further progress on this
issue will require large, multi-field \zs6 galaxy samples with
$J$-band photometry that reaches significantly below \Lchar.

\subsection{Keeping the Universe ionized}\label{ionization}

Recently there has been much interest in the question of the objects
that are responsible for reionizing the Universe at high redshift
(e.g., Lehnert \& Bremer 2003; Bunker et al.\ 2004; Martin \& Sawicki
2004; Stiavelli, Fall, \& Panagia, 2004; Yan \& Windhorst
2004). Several of these authors have concluded that the number of
luminous galaxies at \zs6 is insufficient to provide enough photons to
keep the Universe ionized and have postulated that sub-\Lstar\
galaxies that are largely too faint to be observed directly at these
redshifts may make up the deficit.  The KDF's highest redshift bin is
at \zs4, only 0.7~Gyr after the putative redshift of reionization at
\zeq6.5, and since our survey probes significantly below \Lchar, we 
are in a good position to comment on the issue.  We will study here
whether faint galaxies produce enough ionizing photons to maintain
ionization at \zs4 and later, and ask what that tells us about keeping
the Universe ionized at earlier epochs.

Following much of the \zs6 work (e.g., Ferguson, Dickinson, \&
Papovich 2002; Bunker et al.\ 2004; Yan \& Windhorst 2004; Bouwens et
al.\ 2006), our estimate will be based on the Madau, Haardt, \& Rees
(1999) radiative transfer calculation.  As we mention later, this is
not necessarily the best approach to adopt, but it is the approach
taken by most observers so for consistency we, too, will use it here.

According to Madau, Haardt, \& Rees (1999), the rate of production of
ionizing photons needed to maintain ionization at a given redshift can
be expressed as
\begin{eqnarray}\label{madau.eq}
\dot{\cal N}_{ion}(z) & = & \frac{\bar{n}_H(0)}{\bar{t}_{rec}(z)} \\
	              & = & \Big(\frac{1.58\cdot10^{51}}
			    {\rm s \cdot Mpc^{3}}\Big) 
			    C_{30} \Big(\frac{1+z}{6}\Big)^3 
			    \Big(\frac{\Omega_b h_{70}^2}{0.041}\Big)^2 
			    \nonumber
\end{eqnarray}
where $\bar{n}_H(0)$ is the average density of hydrogen,
$\bar{t}_{rec}(z)$ is the average recombination time, $C_{30}$ is the
clumping factor of ionized hydrogen, $\Omega_b$ is the baryon density,
and $h_{70}$ is Hubble's constant in units of 70 km s\inv Mpc\inv.  It
should be noted that Madau et al.'s $C_{30}$=1, motivated by
simulations by Gnedin \& Ostriker (1997), is likely too high and more
recent studies suggest reionization is somewhat easier to achieve
(see, e.g., Miralda-Escud\'e 2003; Furlanetto \& Oh 2005; Meiksin
2005).  At rest-frame 1700\AA, we are measuring photon energies that
are too low to ionize hydrogen and so we need to relate our 1700\AA\
luminosity densities to number densities of $\lambda$=0--912\AA\
photons.  Using the Starburst99 evolutionary spectral synthesis models
(Leitherer et al.\ 1999) we find that the rate of production of
ionizing photons, $\dot{\cal N}_{ion}$, is related to the 1700\AA\
luminosity $L_{1700}$ by
\begin{equation}\label{lum2rate.eq}
 \dot{\cal N}_{ion}= 1.64 \cdot 10^{25} \frac{L_{1700}}{\rm erg \cdot
 Hz^{-1}}.
\end{equation}
Equation~\ref{lum2rate.eq} is valid for starbursts with age
$>$10$^7$yr, Salpeter (1955) stellar initial mass function (IMF) with
100$>$$M/M_\sun$$>$0.1, and is insensitive to metallicity. Then,
combining Equations~\ref{madau.eq} and \ref{lum2rate.eq} we find that
if the Universe is to be kept ionized by ongoing star formation, we
need to have a 1700\AA\ luminosity density of no less than
\begin{equation}\label{critlumdens.eq}
\frac{L_{1700}^{crit}(z)}{\rm erg \cdot s^{-1} Hz^{-1} Mpc^{-3}} = 
9.63\cdot 10^{25} \frac{C_{30}}{f_{esc}^{rel}}
\Big(\frac{1+z}{6}\Big)^3
\Big(\frac{\Omega_b h_{70}^2}{ 0.041}\Big)^2.
\end{equation}
We have included in Eq.~\ref{critlumdens.eq} a relative escape
fraction term \fesc, which accounts for the fact that in practice, due
to various radiative transfer effects, fewer ionizing photons escape a
galaxy than would be straightforwardly predicted by the number of
escaping {\it non}-ionizing photons.  There is currently much
uncertainty about the escape fraction of ionizing radiation, but it
seems reasonable that \fesc$\lesssim$0.5 (Heckman et al.\ 2001;
Steidel, Pettini, \& Adelberger 2001; Fern\'andez-Soto, Lanzetta, \&
Chen 2003; Inoue et al.\ 2005), and certainly \fesc$\leq$1 since the
relative escape fraction cannot exceed unity.

The amount of UV starlight needed to keep the Universe ionized (as
predicted by Equation~\ref{critlumdens.eq}) is shown as a solid gray
curve in Figs.~\ref{lumdens-bright.fig} and \ref{lumdens-wdust.fig}
for hydrogen clumping factor value of $C_{30}$=1 and a conservative
escape fraction \fesc=0.5. Lower escape fractions would push the curve
upwards while lower clumping factors, $C_{30}$$<$1, would lower it. As
has been pointed out by a number of authors, \zs6 galaxies brighter
than 0.1\Lchar\ do not produce enough UV light to ionize the Universe
(Fig.~\ref{lumdens-bright.fig}).  However, it would seem that there
isn't enough starlight to keep the Universe ionized even at \zs4, and
this is so not only if we restrict ourselves to galaxies brighter than
0.1\Lchar, but even if we integrate down the luminosity function all
the way to $L$=0!

In the calculation presented here, an unevolving \zs4 galaxy
population would clearly not be sufficient to maintain ionization at
\zs6, and yet our observations indicate that the galaxy population is
evolving, with the UV luminosity density dropping with increasing
redshift.  While at \zs4 additional radiation from active galactic
nuclei (AGN) might be enough to make up the deficit (but see Steidel
et al.\ 2001), this is unlikely to be the case at higher redshifts as
the number density of even low-luminosity AGN is decreasing with
increasing lookback time (e.g., Willott et al.\ 2005).  It would thus
seem that the Universe cannot produce enough ionizing radiation either
from stars or black holes to reionize its gaseous content at
\zs6.

There are several ways out of this apparent paradox but ionizing
radiation from very faint but otherwise ``normal'' star-forming
galaxies does not appear to be one of them.  Some possibilities
include a radically non-standard IMF or stars with very low, nearly
primordial metallicities (e.g., Bunker et al.\ 2004; Stiavelli, Fall,
\& Panagia 2004).  A less radical alternative, however, is that the
Madau et al.\ (1999) requirement for reionization that is assumed by
most observers is overly strict.  There is an ongoing debate about the
number of photons needed for reionization, a debate that is related to
the uncertainty in the clumpiness of the intergalactic medium that
controls the importance of recombinations (e.g., Madau et al.\ 1999;
Miralda-Escud\'e 2003; Furlanetto \& Oh, 2005; Meiksin 2005). It now
seems highly plausible that the $C_{30}$=1 assumed by Madau et al.\
(1999) and subsequently by many observers is significantly too high.
If, for example, we follow the results of Meiksin (2005) instead of
those of Madau et al.\ (1999) then the required $L_{1700}^{crit}$ is a
factor of $\sim$5 lower than that shown in
Figs.~\ref{lumdens-bright.fig} and \ref{lumdens-wdust.fig}.  Under
these less restrictive clumping assumptions massive stars in faint
galaxies that follow the even relatively shallow faint-end LF slope we
find in the KDF can produce more than enugh ionizing photons at \zs4
and plausibly also at higher redshifts.  In this case, there is no
need to invoke a very numerous faint population and steep LF faint-end
slope at \zs6.

\subsection{The galaxies that dominate the UV luminosity density at high 
redshift}

The steep increase in the total luminosity density with time at high
redshift is caused by the increase in the number density of
sub-\Lchar\ galaxies and the associated steepening of the LF's
faint-end slope.  However, while the faint end of the luminosity
function evolves, the bright end stays nearly constant --- a fact that
is reflected in the relative flatness of the luminosity density due to
$L$$>$\Lchar\ galaxies.  This differential evolution suggests that
individual galaxies themselves are evolving and that they are doing so
differentially with luminosity.

The evolutionary behavior of the differential luminosity density that
we see over $z$$\sim$5$\rightarrow$0 (Fig.~\ref{lumdens.fig}) is
reminiscent of the galaxy downsizing picture (e.g., Cowie et al.\
1996, Iwata et al.\ 2006a). Here, the epoch of vigorous star formation
in the most luminous galaxies --- those with $L$$>$\Lchar\ --- ends
first with a rapid decline in their activity starting at \zs2.  This
is followed at later epochs by a drop in the activity due to
intermediate-luminosity galaxies ($L$=0.1--1\Lchar).  This in turn is
followed by the dominance of those galaxies in our faintest luminosity
bin --- $L$<0.1\Lchar.  However, unlike in the lower-redshift work of
Cowie et al.\ (1996), here, at higher redshifts, the downsizing
picture is so far only seen in luminosity and it remains to be seen if
it is also tied to galaxy masses.

The differential evolution we observe in the luminosity density and
the luminosity function must reflect evolution of individual galaxies.
However, at present the mechanism for this evolution remains unknown.
In Sawicki \& Thompson (2006) we have suggested several possible
mechanisms related to changes in timescales of star-forming episodes,
or evolution in the amount or distribution of their interestellar
dust.  It is also possible that we are witnessing two different
evolutionary mechanisms that dominate the different ends of the
luminosity function and are competing around the characteristic LF
luminosity \Lstar.  It is possible, for example, that on the two
different sides of \Lstar\ we are seeing galaxies that are governed by
two different regimes of feedback --- e.g., AGN feedback in more
massive (and plausibly more luminous) galaxies, and
star-formation-driven feedback in lower-mass (and lower luminosity)
systems.

We are now working to understand the nature of the evoltion we see by
studying differences in the properties of {\it individual} galaxies as
a function of epoch and luminosity.  Clustering and spectral energy
distribution studies will let us constrain the dark halo and stellar
masses of galaxies as a function of luminosity, and thus will tell us
whether the downsizing we see is related to galaxy masses as it seems
to be at lower redshifts (Cowie et al.\ 1996).  SED studies will also
let us test whether other galaxy properties, such as their dust
properties or timescales of their starbursting episodes vary with
luminosity and epoch and thus whether they are responsible for the
evolution we see.  While observationally expensive, such differential
comparisons of faint members of a population with their brighter
brethren can lead to important insights (see, e.g., Sawicki et al.\
2005) and will be key to understanding what drives the evolution of
galaxies at high redshift.

\section{SUMMARY AND CONCLUSIONS}

Our KDF survey has allowed us to make the best measurement thus far of
the UV luminosity density at \zs4, 3, and 2.  This is the case because
our work builds directly on the well-tested photometric selection
techniques of Steidel et al.\ (1999, 2003, 2004), because it probes
very deep into the galaxy luminosity function, and because it does so
using several spatially independent fields to control for the effects
of cosmic variance.  The main results of our analysis of the UV
luminosity density are:

\begin{enumerate}

\item The picture of {\it total} UV luminosity density, i.e., luminosity 
density due to galaxies of all luminosities, is that of a rise from
early epochs, $z$$\geq$4, followed by either a plateau or a broad peak
at \zs3--1, and then followed by a decline to \zeq0.  This picture is
consistent with that presented by Sawicki et al.\ (1997), although is
at odds with some recent work that posits a constant luminosity
density over \zs6--1.

\item The luminosity density at high redshift, \zs2--4, is dominated 
not by luminous galaxies, but by sub-\Lchar\ galaxies, where \Lchar\
is the characteristic 1700\AA\ luminosity that corresponds to
\Mchar= $-21.0$.  Indeed, it is galaxies within a rather narrow
luminosity range $L$=0.1--1\Lchar\ that dominate at these redshifts
and it is the rapid evolution of these galaxies that drives the
evolution in the {\it total} UV luminosity density.

\item In contrast to $z$$\gtrsim$2, at lower redshifts, $z$$\lesssim$1 
the total UV luminosity density, as shown most recently by GALEX, is
dominated by {\it very} faint galaxies with $L$$<$0.1\Lchar. The
$L$=0.1--1\Lchar\ galaxies that dominated at $z$$\gtrsim$2 are still
important but no longer dominant.  This reversal in the luminosity of
the galaxies that dominate the total luminosity density is reminiscent
of galaxy downsizing, although the effect we see so far is a
downsizing in UV luminosity rather than galaxy mass.

\item Several recent studies have suggested that at \zs6 there are not
enough luminous galaxies to maintain ionization of intergalactic gas
and proposed that a large population of faint galaxies is needed for
that task.  Within the context of the most widely-used models, we find
that there seemingly aren't enough bright {\it and} faint star-forming
galaxies even as late as \zs4 to maintain ionization, questioning the
assumptions of these most widely-used models.  Apparently the Universe
must be easier to reionize than some recent studies have assumed.

\end{enumerate}

In this paper we have shown that the total UV luminosity density of
the Universe at high redshift is dominated by galaxies fainter than
the \Lchar\ galaxies that have thus far drawn most of the community's
attention.  These sub-\Lchar\ galaxies are, as we have also found in
Sawicki \& Thompson (2006), evolving rapidly as a population, while
the population of their far better studied, more luminous cousins
appears to remain unchanged with time.  The question then becomes,
what makes the high-$z$ galaxies brighter than \Lchar\ and those
fainter than \Lchar\ so different.  We will address this issue in the
future by studying the clustering properties and stellar populations
of these important systems.


\vspace{5mm}

We thank the time allocation committee for a generous time allocation
that made this project possible and the staff of the W.M.\ Keck
Observatory for their help in obtaining these data.  We are also
grateful to Chuck Steidel for his encouragement and support of the KDF
project, Jerzy Sawicki for many useful comments, Ikuru Iwata, Kouji
Ohta, and their collaborators for sharing their new \zs5 Subaru
results in advance of publication and Crystal Martin and Peng Oh for
useful discussions.  Finally, we wish to recognize and acknowledge the
very significant cultural role and reverence that the summit of Mauna
Kea has always had within the indigenous Hawaiian community; we are
most fortunate to have the opportunity to conduct observations from
this mountain.

\newpage



\clearpage




\clearpage

\begin{deluxetable}{lccccc}
\tablewidth{0pt} 
\tablecaption{\label{lumdens.tab} Rest-frame 1700\AA\ UV luminosity densities\tablenotemark{a}}
\tablehead{
\colhead{$z$} & 
\colhead{total} & 
\colhead{$L$$>$\Lchar} & 
\colhead{1$>$$L/$\Lchar$>$0.1} & 
\colhead{$L$$<$0.1\Lchar} & 
\colhead{survey}
}
\startdata
0.05&  25.56$^{+0.08}_{-0.08}$ & 18.56$^{+8.45}_{-8.45}$ &  24.74 $^{+0.14}_{-0.14}$ &  25.49 $^{+0.08}_{-0.08}$ & GALEX\\  
0.3 &  25.84$^{+0.17}_{-0.17}$ & 20.73$^{+4.90}_{-4.90}$ &  25.24 $^{+0.25}_{-0.25}$ &  25.72 $^{+0.17}_{-0.17}$ & GALEX\\  
0.5 &  25.96$^{+0.40}_{-0.40}$ & 23.54$^{+0.83}_{-0.83}$ &  25.48 $^{+0.29}_{-0.29}$ &  25.78 $^{+0.47}_{-0.47}$ & GALEX\\  
0.7 &  26.14$^{+0.56}_{-0.56}$ & 24.19$^{+0.81}_{-0.81}$ &  25.69 $^{+0.32}_{-0.32}$ &  25.94 $^{+0.66}_{-0.66}$ & GALEX\\  
1.0 &  26.12$^{+0.78}_{-0.78}$ & 24.45$^{+0.80}_{-0.80}$ &  25.69 $^{+0.33}_{-0.33}$ &  25.91 $^{+0.97}_{-0.97}$ & GALEX\\  
1.7 &  26.75$^{+0.02}_{-0.02}$ & 25.59$^{+0.15}_{-0.15}$ &  26.63 $^{+0.02}_{-0.02}$ &  26.03 $^{+0.05}_{-0.05}$ & KDF\\  
2.2 &  26.43$^{+0.03}_{-0.03}$ & 25.66$^{+0.09}_{-0.09}$ &  26.21 $^{+0.02}_{-0.02}$ &  25.76 $^{+0.10}_{-0.10}$ & KDF\\  
3.0 &  26.42$^{+0.03}_{-0.03}$ & 25.64$^{+0.03}_{-0.03}$ &  26.15 $^{+0.02}_{-0.02}$ &  25.91 $^{+0.10}_{-0.10}$ & KDF\\  
4.0 &  26.06$^{+0.07}_{-0.07}$ & 25.47$^{+0.05}_{-0.05}$ &  25.81 $^{+0.05}_{-0.05}$ &  25.34 $^{+0.25}_{-0.25}$ & KDF\\  
4.8 &  25.86$^{+0.09}_{-0.09}$ & 25.33$^{+0.08}_{-0.08}$ &  25.63 $^{+0.09}_{-0.09}$ &  24.93 $^{+0.33}_{-0.33}$ & Iwata et al.\ (2006)\tablenotemark{b}\\  
\enddata
\tablenotetext{a}{In logarithmic units of erg/s/Hz/Mpc$^3$}
\tablenotetext{b}{Computed by us from data kindly provided by these authors in advance of publication}
\end{deluxetable}

\begin{deluxetable}{lccccc}
\tablewidth{0pt} 
\tablecaption{\label{lumdensBM.tab} UV luminosity densities in galaxies brighter than 0.1\Lchar}
\tablehead{
\colhead{Survey} & 
\colhead{$z$} & 
\colhead{$\log (\rho_{UV}\tablenotemark{a})$} 
}
\startdata
GALEX& 						0.05    & $24.74 ^{+0.14}_{-0.14} $ \\
	& 					0.3    &  $25.24 ^{+0.25}_{-0.25} $ \\
	& 					0.5    &  $25.48 ^{+0.29}_{-0.29} $ \\
	& 					0.7    &  $25.70 ^{+0.34}_{-0.34} $ \\
	& 					1.0    &  $25.71 ^{+0.35}_{-0.35} $ \\
KDF& 						2.2    &  $26.32 ^{+0.02}_{-0.02} $ \\
	& 					3.0    &  $26.26 ^{+0.01}_{-0.01} $ \\
	& 					4.0    &  $25.97 ^{+0.03}_{-0.03} $ \\
Iwata et al.\ (2006)\tablenotemark{b}& 		4.8     & $25.81 ^{+0.07}_{-0.07} $ \\
Steidel et al.\ (1999) & 			3	& $26.30 ^{+0.06}_{-0.06} $ \\
			& 			4	& $26.22 ^{+0.10}_{-0.10} $ \\
Giavalisco et al.\ (2004) \chisq\ method& 	3.78	& $26.33 ^{+0.05}_{-0.05} $ \\
					& 	4.92	& $26.14 ^{+0.08}_{-0.08} $ \\
					& 	5.74	& $26.18 ^{+0.24}_{-0.19} $ \\
Giavalisco et al.\ (2004) \Veff\ method& 	3.78	& $26.12 ^{+0.05}_{-0.05} $ \\
					& 	4.92	& $25.85 ^{+0.08}_{-0.08} $ \\
					& 	5.74	& $24.92 ^{+0.24}_{-0.19} $ \\
Bunker et al.\ (2004)& 				6	& $25.55 ^{+0.06}_{-0.06} $ \\
Bouwens et al.\ (2006)& 			6	& $26.10 ^{+0.05}_{-0.05} $ \\
Ouchi et al.\ (2004)& 				4.0	& $26.38 ^{+0.04}_{-0.04} $ \\
		& 				4.7	& $26.20 ^{+0.16}_{-0.16} $ \\
Gabasch et al.\ (2004a, 2004b)& 		2.0	& $26.21 ^{+0.04}_{-0.04} $ \\
				& 		2.7	& $26.28 ^{+0.05}_{-0.05} $ \\
				& 		3.5	& $26.25 ^{+0.02}_{-0.02} $ \\
				& 		4.5	& $25.96 ^{+0.04}_{-0.04} $ \\
\enddata
\tablenotetext{a}{Rest-1700\AA\ luminosity density in units of erg/s/Hz/Mpc$ ^3$}
\tablenotetext{b}{Computed by us from data kindly provided by these authors in advance of publication}
\end{deluxetable}


\begin{thebibliography}{}

\bibitem[Arfken (1985)]{arf85} Arfken, G. 1985, Mathematical Methods for 
Physicists (San Diego: Academic Press, Inc.)

\bibitem[Arnouts et al.\ (2005)]{arn05} Arnouts, S.\ et al.\ 2005, \apjl, 
619, L43

\bibitem[Baker et al.\ (2001)]{bak01} Baker, A. J., Lutz, D., Genzel, R., 
Tacconi, L. J., Lehnert, M. D. 2001, \aap, 372, 37

\bibitem[Barger et al.\ (1998)]{bar98} Barger, A.J., et al.\ 1998, 
\nat, 394, 248

\bibitem[Bouwens et al.\ (2003)]{bou03} Bouwens, R.J., et al.\ 2003, \apj, 
595, 589 

\bibitem[Bouwens et al.\ (2004)]{bou04} Bouwens, R.J., et al.\ 2004, \apj, 
606, L25 

\bibitem[Bouwens et al.\ (2006)]{bou06} Bouwens, R.J., Illingworth, G.D., 
Blakeslee, J.P., \& Franx, M. 2006, \apj, submitted, astro-ph/0509641

\bibitem[Bunker et al.\ (2004)]{bun04} Bunker, A.J., Stanway, E.R., Ellis, 
R.S., McMahon, R.G. 2004, \mnras, 355, 374 

\bibitem[Chapman et al.\ (2000)]{cha00} Chapman, S.C., Scott, D., 
Steidel, C.C., Borys, C., Halpern, M., Morris, S.L., Adelberger, K.L., 
Dickinson, M., Giavalisco, M., \& Pettini, M. 2000, \mnras, 319, 318

\bibitem[Chapman et al.\ (2005)]{cha05} Chapman, S.C., Blain, A.W., 
Smail, I., \& Ivison, R.J. 2005, \apj, 622, 772

\bibitem[Connolly et al.\ (1997)]{con97} Connolly, A.J., Szalay, A.S., 
Dickinson, M., SubbaRao, M.U., \& Brunner, R.J. 1997, \apjl, 486, L11

\bibitem[Cowie et al.\ (1996)]{cow96} Cowie, L.L., Songaila, A., 
Hu, E.M., \& Cohen, J.G. 1996, \aj, 112, 839


\bibitem[Eales et al.\ (1999)]{eal01} Eales, S., et al.\ 1999, \apj, 515, 518

\bibitem[Ferguson et al.\ (2000)]{fer00} Ferguson, H.C., Dickinson, M., \& 
Williams, R. 2000, \araa, 38, 667

\bibitem[Ferguson et al.\ (2002)]{fer02} Ferguson, H.C., Dickinson, M., 
\& Papovich, C. 2002, \apj, 569, L65

\bibitem[Fern\'andez-Soto et al.\ (2003)]{fer03} Fern\'andez-Soto, A., 
Lanzetta, K.M., \& Chen, H.-W. 2003, \mnras, 342, 1215

\bibitem[Gabasch et al.\ (2004a)]{gab04a} Gabasch, A. et al.\ 2004a, 
\aap, 421, 41 

\bibitem[Gabasch et al.\ (2004b)]{gab04b} Gabasch, A. et al.\ 2004b, \apjl,
616, L83 

\bibitem[Giavalisco et al.\ (2004a)]{gia04a} Giavalisco, M. et al.\ 2004a, 
\apjl, 600, L93 

\bibitem[Giavalisco et al.\ (2004b)]{gia04b} Giavalisco, M. et al.\ 2004b, 
\apjl, 600, L103 

\bibitem[Gnedin \& Ostriker (1997)]{gne97} Gnedin, N.Y. \& Ostriker, J.P. 
1997, \apj, 486, 581

\bibitem[Hall et al.\ (2001)]{hal01} Hall, P.B., Sawicki, M., Martini, P.,
Finn, R.A., Pritchet, C.J., Osmer, P.S., McCarthy, D.W., Evans, A.S., 
Lin, H., \& Hartwick, F.D.A. 2001, \aj, 121, 1840

\bibitem[Heckman et al.\ (2001)]{hec01} Heckman, T.M., Sembach, K.R., 
Meurer, G.R., Leitherer, C., Calzettin, D., \& Martin, C.L. 2001,
\apj, 558, 56

\bibitem[Inoue et al.\ (2005)]{ino05} Inoue, A.K., Iwata, I., Deharveng, 
J.-M., Buat, V., \& Burgarella, D. 2005, \aap, 435, 471

\bibitem[Iwata et al.\ (2003)]{iwa03} Iwata, I., Ohta, K., Tamura, N., 
Ando, M., Wada, S., Watanabe, C., Akiyama, M., \& Aoki, K. 2003,
\pasj, 54, 415

\bibitem[Iwata et al.\(2005)]{iwa05} Iwata, I., Inoue, A.K., \& 
Burgarella, D. 2005, \aap, 440, 881

\bibitem[Iwata et al.\ (2006a)]{iwa06a} Iwata, I., Ohta, K., Ando, M., 
Kiuchi, G., Tamura, N., Akiyama, M., \& Aoki, K. 2006a,
astro-ph/0510829

\bibitem[Iwata et al.\ (2006b)]{iwa06b} Iwata, I., et al.\ 2006b, 
in preparation

\bibitem[Kennicutt et al.\ (1998)]{ken98} Kennicutt Jr., R.C., 1998, \araa, 
36, 189 

\bibitem[Lehnert \& Bremmer (2003)]{leh03} Lehnert, M.D., Bremer, M.N. 2003, 
\apj, 593, 630

\bibitem[Leitherer et al.\ (1999)]{lei99} Leitherer, C., Schaerer, D., 
Goldader, J.D., Gonz\'alez Delgado, R.M., Robert, C., Kune, D.F., de
Mello, D.F., Devost, D., Heckman, T. 1999, \apjs, 123, 3

\bibitem[Lilly et al.\ (1996)]{lil96} Lilly, S.J., Le F\'evre, O., 
Hammer, F., \& Crampton, D. 1996, \apjl, 460, L1

\bibitem[Madau (1995)]{mad95} Madau, P. 1995, \apj, 441, 18

\bibitem[Madau et al.\ (1996)]{mad96} Madau, P., Ferguson, H.C., 
Dickinson, M.E., Giavalisco, M., Steidel, C.C., \& Fruchter, A. 1996,
\mnras, 283, 1388

\bibitem[Madau et al.\ (1998)]{mad98} Madau, P., Pozetti, L., \& Dickinson, M. 
1998, \apj, 498, 106

\bibitem[Madau et al.\ (1999)]{mad99} Madau, P., Haardt, F., \& Rees, M.J. 1999, 
\apj, 514, 648

\bibitem[Martin \& Sawicki (2004)]{mar04} Martin, C.L. \& Sawicki, M. 2004, 
\apj, 603, 414

\bibitem[Meiksin (2005)]{mei05} Meiksin, A. 2005, \mnras, 356, 596

\bibitem[Miralda-Escud\'e (2003)]{mir03} Miralda-Escud\'e, J. 
2003, \apj, 597, 66

\bibitem[Furlanetto \& Oh (2005)]{fur03} Furlanetto, S.R.\ \& Oh, S.P. 2005, 
\mnras, 363, 1031

\bibitem[Oke (1974)]{oke74} Oke, J.B., 1974, \apjs, 236, 27

\bibitem[Ouchi et al.\ (2004)]{ouc04a} Ouchi, M., et al. 2004, \apj, 
611, 660

\bibitem[Papovich et al.\ (2001)]{pap01} Papovich, C., Dickinson, M., \& 
Ferguson, H.C. 2001, \apj, 559, 620

\bibitem[P\'erez-Gonzalez et al.\ (2003)]{per03} P\'erez-Gonz\'alez, P.G., 
Zamorano, J., Gallego, J., Arag\'on-Salamanca, A., Gil de Paz,
A. 2003, \apj, 591, 827

\bibitem[P\'erez-Gonzalez et al.\ (2005)]{per05} P\'erez-Gonz\'alez, P.G.
et al.\ 2005, \apj, 630, 82

\bibitem[Press et al.\ (1986)]{pre86} Press, W.H., Flannery, B.P., 
Teukolsky, S.A., \& Vetterling, W.T. 1986, Numerical Recipes
(Cambridge: Cambridge University Press)

\bibitem[Reddy \& Steidel (2004)]{red04} Reddy, N.A. \& Steidel, C.C. 2004, 
\apjl, 603, L13

\bibitem[Reddy et al.\ (2005)]{red05} Reddy, N.A., Erb, D.K., Steidel, C.C., 
Shapley, A.E., Adelberger, K.L., \& Pettini, M. 2005, \apj, 633, 748

\bibitem[Salpeter (1955)]{sal55} Salpeter, E.E. 1955, \apj, 121, 161

\bibitem[Sawicki et al.\ (1997)]{saw97} Sawicki, M.J., Lin, H., \& Yee,
H.K.C. 1997, \aj, 113, 1 

\bibitem[Sawicki \& Yee (1998)]{saw98} Sawicki, M. \& Yee, H.K.C. 1998, 
\aj, 115, 1329  

\bibitem[Sawicki (2001)]{saw01} Sawicki, M. 2001, \aj, 121, 2405 

\bibitem[Sawicki et al.\ (2005)]{saw05a} Sawicki, M., Stevenson, M., 
Barrientos, L.F., Gladman, B., Mall\'en-Ornelas, G., \& van den Bergh,
S. 2005, \apj, 627, 621 

\bibitem[Sawicki \& Thompson (2005)]{saw05} Sawicki, M. \& Thompson, D. 2005,
\apj, 635, 100 (\kdfdata)

\bibitem[Sawicki \& Thompson (2006)]{saw06a} Sawicki, M. \& Thompson, D. 2006,
\apj, 642, 653, (\kdflf)

\bibitem[Sawicki \& Webb (2005)]{saw05c} Sawicki, M. \& Webb, T.M.A. 2005, 
\apjl, 618, L67

\bibitem[Sawicki et al.\ (2006)]{saw06b} Sawicki, M.\ et al.\ 2006, in prep.

\bibitem[Schechter et al.\ (1976)]{sch76} Schechter, P. 1976, \apj, 203, 297

\bibitem[Schiminovich et al.\ (2005)]{sch05} Schiminovich, D., et al.\ 2005, 
\apjl, 619, L47

\bibitem[Shapley et al.\ (2001)]{sha01} Shapley, A.E., Steidel, C.C., 
Adelberger, K.L., \& Pettini, M. 2001, \apj, 562, 95

\bibitem[Shapley et al.\ (2005)]{sha05} Shapley, A.E., Steidel, C.C., 
Erb, D.K., Reddy, N.A., Adelberger, K.L., Pettini, M., Barmby, P., \& 
Huang, J. 2005, \apj, 626, 698

\bibitem[Smail et al.\ (1997)]{sma97} Smail, I., Ivison, R.J., 
\& Blain, A.W. 1997, \apj, 490, L5

\bibitem[Steidel et al.\ (1999)]{ste99} Steidel, C.C., Adelberger,
K.L., Giavalisco, M., Dickinson, M., \& Pettini, M. 1999, \apj, 519, 1

\bibitem[Steidel et al.\ (2001)]{ste01} Steidel, C.C., Pettini, M., 
\& Adelberger, K.L. 2001, \apj, 546, 665

\bibitem[Steidel et al.\ (2003)]{ste03} Steidel, C.C., Adelberger, K.L., 
Shapley, A.E., Pettini, M., Dickinson, M., \& Giavalisco, M. 2003,
\apj, 592, 728

\bibitem[Steidel et al.\ (2004)]{ste04} Steidel, C.C., Shapley, A.E., 
Pettini, M., Adelberger, K.L., Erb, D.K., Reddy, N.A., \& Hunt, M.P, 
2004, \apj, 604, 534

\bibitem[Stiavelli et al.\ (2004)]{sti04} Stiavelli, M., Fall, S.M., 
\& Panagia, N. 2004, \apj, 610, L1 

\bibitem[Tresse et al.\ 2002]{tre02} Tresse, L., Maddox, S.J., 
Le F\`evre, O., \& Cuby, J.-G. 2002, \mnras, 337, 369

\bibitem[Vijh, Witt, \& Gordon (2003)]{vij03} Vijh, U., Witt, A.N., 
\& Gordon, K.D. 2003, \apj, 587, 533

\bibitem[Webb et al.\ (2003)]{web03} Webb, T.M.A., Eales, S., Foucaud, S., 
Lilly, S. J., McCracken, H., Adelberger, K., Steidel, C., Shapley, A.,
Clements, D. L., Dunne, L., Le F\'evre, O., Brodwin, M., Gear,
W. 2003, \apj, 582, 6

\bibitem[Williams et al.\ (1996)]{wil96}Williams, R.J., et al.\ 1996, 
\aj, 112, 1335

\bibitem[Willott et al.\ (2005)]{wil05}Willott, C.J., Delfosse, X., 
Forveille, T., Delorme, P., \& Gwyn, S. 2005, \apj, 633, 630

\bibitem[Wyder et al.\ (2005)]{wyd05} Wyder, T.K., et al.\ 2005, \apjl, 
619, L15

\bibitem[Yan \& Windhorst (2004)]{yan04}Yan, H. \& Windhorst, R.A. 2004, 
\apjl, 600, L1 

\end{thebibliography}
\end{document}